\documentclass[runningheads]{llncs}
\usepackage{graphicx}
\usepackage{times}
\usepackage{helvet}
\usepackage{courier}

\usepackage{tabularx}
\usepackage{dcolumn}
\usepackage{booktabs}
\usepackage{url}
\usepackage{subfigure}
\usepackage{balance}  
\usepackage{graphics} 
\usepackage[usenames,dvipsnames]{xcolor}
\usepackage{graphicx}
\usepackage{dcolumn}
\usepackage{enumitem}
\newcounter{JSNumberOfComments}
\stepcounter{JSNumberOfComments}

\newcommand{\am}{\texttt{Asian male}}
\newcommand{\af}{\texttt{Asian female}}

\newcommand{\km}{\texttt{Black male}}
\newcommand{\kf}{\texttt{Black female}}

\newcommand{\wm}{\texttt{White male}}
\newcommand{\wf}{\texttt{White female}}

\newcommand{\asian}{\texttt{Asian}}

\newcommand{\black}{\texttt{Black}}

\newcommand{\white}{\texttt{White}}

\newcommand{\female}{\texttt{Female}}
\newcommand{\male}{\texttt{Male}}

\begin{document}
\title{Gender and Racial Diversity in Commercial Brands' Advertising Images on Social Media}
\subtitle{[Please cite the SocInfo'19 version of this paper]}
\titlerunning{Gender and Racial Diversity in Advertising Images on Social Media}
%
\author{Jisun An\inst{1}\orcidID{0000-0002-4353-8009} \and
Haewoon Kwak\inst{1}\orcidID{0000-0003-1418-0834}
\authorrunning{J. An and H. Kwak}
\institute{Qatar Computing Research Institute, Hamad Bin Khalifa University\\
\email{\{jisun.an,haewoon\}@acm.org}}
}

\maketitle              
\begin{abstract}
Gender and racial diversity in the mediated images from the media shape our perception of different demographic groups. 
In this work, we investigate gender and racial diversity of 85,957 advertising images shared by the 73 top international brands on Instagram and Facebook. 
We hope that our analyses give guidelines on how to build a fully automated watchdog for gender and racial diversity in online advertisements.
\end{abstract}

\section{Introduction}

``The \textit{perception of the other} is a core aspect of the integration of ethnic minorities and immigrants~\cite{trebbe2011ethnic}.'' 
Integration and cooperation in a multicultural society can be easy or challenging, depending on the perception of self and other groups. 
One of the notable channels that affect people's perception of the other is mass media~\cite{gerbner2002growing,kwak2018we}. 
After being repeatedly exposed, viewers' belief and attitudes are shaped by the mediated images from the media. 
Thus, the portrayal of different demographic groups in media, particularly advertisements, can play an essential role in shaping the formation of identities of those groups~\cite{choudhury1974black}.

For the decades, scholars have studied the portrayals of demographic groups (mainly ethnic minorities compared to Whites) in advertisements in print and broadcast media~\cite{shuey1953stereotyping,stevenson2007six}. 
Minority groups have been found to be continuously under- or misrepresented in the advertisements. 
For example, advertisements perpetuate conventional stereotyped images of races and gender, which are, for instance, African Americans depicted as aggressive and active, and women portrayed as young, thin, and smiling~\cite{coltrane2000perpetuation}. 
Despite the warnings from these studies, controversial advertisements, such as the one by Hornbach~\cite{hornbach}, are still being created. Thus, the continuous monitoring of advertisements and raising public awareness are essential.

In monitoring advertisements, analyzing the visual content is one of the main components because most advertisements include visual elements because of their effectiveness in marketing~\cite{mcquarrie2008differentiating}.
However, previous studies on examining advertisements largely depend on manual efforts for analyzing the visual content~\cite{shuey1953stereotyping,bailey2006year,shabbir2014deconstructing}.
In the modern world, which is overloaded by digital content published on the web and social media, how to minimize such manual efforts and build an automated solution is the key to monitoring the tremendous volume of advertisements.

In this work, as a first step, we propose an automated way to examine gender and racial diversity in online advertisements in a large scale using modern image analysis techniques. 
Through a comprehensive review of previous literature, we define three metrics of the gender and racial diversity in advertisements that can be computed by automated tools: (1) how many times each gender and race appear, (2) how many times each gender and race appear in cross-sex interaction context, and 
(3) how many times each gender and race appear as smiling faces.

Using the three metrics, we demonstrate a large-scale analysis of the gender and racial diversity in the 85,957 advertising images of 73 international brands on Instagram and Facebook. 
We define advertising images of brands in a broad sense: as images posted on social media by their official accounts. Considering the definition of advertisement, which is ``a picture, sign, etc. that is used to make a product or service known and persuade people to buy it,'' it is reasonable to consider the images posted on social media by brands' official accounts as their advertising images.
As the marketing in social media has been differentiating from that in traditional media~\cite{hanna2011we}, our analysis of advertising images on social media adds new dimensions to the stream of previous research on U.S.-based empirical studies of advertisements in mass media~\cite{shuey1953stereotyping,cox1969changes,bailey2006year}.
In particular, the following research questions drawn from previous literature guide our analysis: 
\begin{itemize}
    \item[] RQ1: How many times does each demographic group appear in advertising images? How different are they across the brands? 
    \item[] RQ2: Which pairs of demographic groups are preferred in advertising images for cross-sex interaction context?
    \item[] RQ3: Is there a specific demographic group depicted more with smiling faces?
\end{itemize}

Our study provides a holistic view of the gender and racial diversity in today's advertising images by global brands on social media and is a great demonstration of the feasibility of our proposed metrics computed in an automated way.

\section{Related Work}

For decades there has been a long research stream on depictions of different demographic groups in marketing communications including advertisements~\cite{shuey1953stereotyping,cox1969changes,bailey2006year,shabbir2014deconstructing,coltrane2000perpetuation}.
Both cross-sectional and longitudinal U.S.-based studies have looked into the racial mix of models in TV commercials and print advertisements~\cite{millard2006stereotypes,taylor1994not,bowen1997minority}.  

While some variations exist across the different data sources~\cite{plous1997racial,bailey2006year}, a general consensus on depictions of ethnic minorities in advertisements, African American models in particular, has emerged: 
(1) increasing appearances, even compared to the incidence of their population relative to the entire U.S. population~\cite{stevenson2007six}, 
(2) decreasing portrayals of blue-collar workers and increasing portrayals of professionals~\cite{zinkhan1990use}, and 
(3) taking more major roles~\cite{lee2005portrayal}.  
Along with studies on African American models, representations of Asians and Hispanics in advertisements have also been investigated~\cite{paek2003racial,bowen1997minority,lee2005portrayal}. 
Compared to other demographic groups, Asian American models are frequently portrayed as a work-centric `model minority'~\cite{taylor1997asian}, while Hispanic models are still underrepresented compared to their populations~\cite{paek2003racial}. 
Nonetheless, the images of `White hegemonic masculinity and White feminine romantic fulfillment' are perpetuated in virtually all forms of television marketing~\cite{coltrane2000perpetuation}.

Such racial stereotypes determine the roles of models and the types of products.
For example, Black models are frequently depicted in sports/athletic contexts and thus are overrepresented in advertisements of sports/athletic products~\cite{mastro2003representations}. 
As Asian models are excessively depicted in work settings, they tend to appear in advertisements associated with affluence or work life and not to appear in advertisements of home- or social life-related products~\cite{taylor1997asian}. 
By contrast, White models are depicted more in home settings or cross-sex interactions because their stereotypes have both masculinity and feminine. As a result, they are overrepresented in the advertisements for upscale, beauty, and home-related products~\cite{jacobs2003race}.

The impact of these stereotyped- or biased advertisements has been extensively discussed. 
Simply, those advertisements make people hold a biased belief by reinforcing negative stereotypes~\cite{cohen2004advertisements}. 
In communication studies, two theories have offered a theoretical foundation for a better explanation: 
social learning theory~\cite{bandura1969social} and cultivation theory~\cite{gerbner2002growing}.
Both theories explain how our perception can be influenced by  the depictions of different demographic groups in advertisements. Social learning theory posits that we can learn general behaviors and attitudes through observing others instead of through first-hand experience. 
Similarly, watching stereotyped depictions in advertisements are known to have a negative impact on people's perceptions~\cite{mccullick2003butches,skill1990family}. 
On the other hand, cultivation theory suggests viewers' beliefs and attitudes are shaped by the mediated images to which they are repeatedly exposed. 
Particularly, our perception is likely to be cultivated when our social reality is close to the mediated images in advertisements~\cite{choudhury1974black}. For example, Asian viewers' perception is likely to be cultivated when they see Asian models in advertisements. 

Subliminal effects of depictions in advertisements have also been studied~\cite{katz1988racial}. They are typically tested by the Implicit Association Test~\cite{greenwald1998measuring}, measuring response latencies to a given stimulus.  For example, by comparing the response latencies to good/bad words associated with images of the faces of the same and different races, implicit in-group preferences are observed~\cite{richeson2005brief}.

\section{Data Collection}

We begin by compiling a list of global brands and their sectors (e.g., apparel, beverage, and so on) by merging lists of global brands in Brand Finance's Global 500 2016~\cite{brandirectory} and Interbrand's Best Global Brands 2016~\cite{interbrand}.
We then refine the list of the brands by using the following criteria: First, as business-to-business (B2B) and business-to-consumer (B2C) brands exhibit significant differences in their social media usage~\cite{swani2014should}, in this work, we focus only on B2C brands, which directly face users in their business. 
Second, to mitigate differences in cultures and attitudes toward diversity, we consider only the brands that originated from the U.S.
We then filter out the brands that do not have an official Instagram or Facebook account.
We note that these brands tend to have multiple Instagram or Facebook accounts targeting other regions (e.g., Starbucks Middle East) as they are international brands. In that case, we consider only the primary account.
We then use Facebook Graph API and Instagram API to collect all the images uploaded till November 2017 based on the list of Instagram and Facebook accounts. 
After collecting the data, we eliminate the brands that do not have more than 50 images with faces. Finally, we have 73 international brands and their 85,957 advertising images posted on Instagram and Facebook.

To infer demographic attributes of models in advertisements automatically, we use Face++~\cite{facepp}, which is one of the widely used commercial tools for face detection. Face++ has been popularly used in computational social science research, showing reliable performance in inferring gender (Male or Female) and race (White, Asian, or Black)~\cite{zagheni2014inferring,an2016greysanatomy,reis2017demographics,an2018diversity}. 
In a recent study~\cite{jung2018facepp}, Face++ achieved more than 90\% accuracy in gender and race detection and performed better than or as good as other commercial tools across diverse datasets.
We note that we do not use the age inference of Face++ because the reliability of age inference has not been widely tested.
Face++ also detects smiling levels (the degree of smile) of each face, which is used for answering RQ3. 
We do not consider images that Face++ returns an inference with low confidence scores ($<0.7$).

In the remaining part of this paper, we use the \texttt{typewriter} font to refer to a demographic group inferred by Face++. For example, we use \asian~or \km, to shorten the phrase `inferred as Asian by Face++' or `inferred as Black male by Face++' for brevity.

\section{Frequencies of Appearances of Each Demographic Group}

Along with a line of previous studies~\cite{taylor1997asian,mastro2003representations,cox1969changes,coltrane2000perpetuation}, we start with examining frequencies of appearances of each demographic group in advertising images on social media. 
Each brand might have different strategies to use models for the best promotion of their products and services. 
Accordingly, our 73 brands across different sectors might have a wide range of variations in the proportions of each demographic group. 

\begin{figure}[ht!]
 \begin{center}
 \includegraphics[width=0.8\columnwidth]{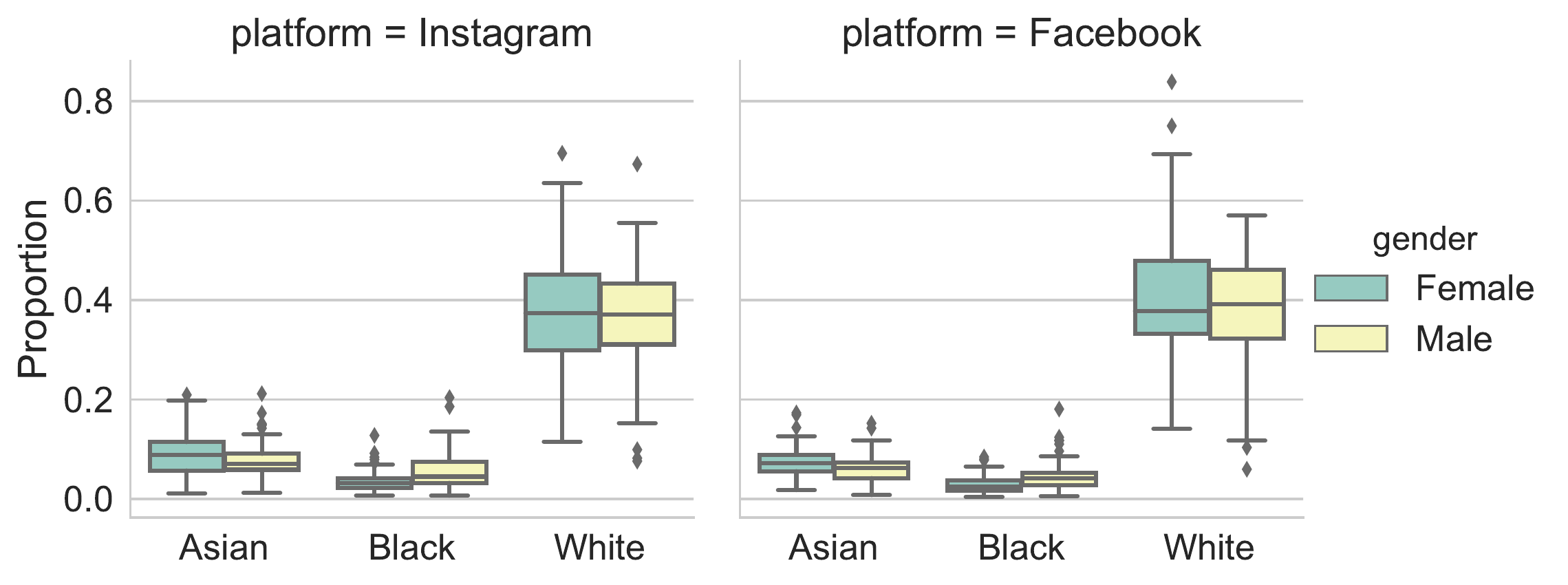}
 \caption{Proportions of appearances of each demographic group in advertising images} 
 \label{fig:simple_appearance}
 \end{center}
\end{figure}

Figure~\ref{fig:simple_appearance} shows the proportion of appearances of each demographic group that appeared in advertising images on Instagram and Facebook. 
While there are some differences between the platforms, we observe consistent trends. 
Notably, appearances of \wf~(Mean: 0.38 on Instagram, and 0.41 on Facebook) and \wm~models (Mean: 0.36 on Instagram, and 0.38 on Facebook) outnumber the other demographic groups on both Instagram and Facebook. 
These results are statistically significant as confirmed by Kruskal-Wallis 1-way ANOVA test with Dunn's post-hoc test ($H$=324.9 for Instagram and 331.9 for Facebook and all the adjusted $p$ from Dunn's test are lower than 2.34e-12).
By contrast, \kf~and \km~models show a considerably lower number of appearances. 
The percentage of appearances of \kf~models is strikingly low, which is on average 2.9\% and 3.6\% of models on Instagram and Facebook, respectively. Considering the recent trends in which the percentage of Black female models is increasing in other media~\cite{shabbir2014deconstructing}, they are quite underrepresented in advertising images on social media. 

Compared with the actual population, the percentage of the appearances of each racial group in advertising images is enticing our attention. 
In the 2010 Census~\cite{uscensus}, Asian, Black, and White groups account for 5\%, 13\%, and 72\% of the U.S. population, respectively. As the percentages in the census are computed with other racial groups (e.g., American Indian) as well, for a fair comparison, the percentage of each race should be normalized by considering the three races only, which are 5.6\%, 14.4\%, and 80\% for Asian, Black, and White groups, respectively. 
In the advertising images on social media, we find that \asian, \black, and \white~models account for 16.8\%, 8.9\%, and 74.2\% on Instagram and 13.3\%, 7.4\%, and 79.3\% on Facebook, respectively.

We observe that the percentage of Whites in the actual population and those that appeared in social media advertisements are comparable. 
In contrast, \black~models appear much less in social media advertisements considering their actual population.
Surprisingly, Asians are overrepresented almost three times relative to their population in the U.S. 
Our results on social media show some commonality and contrasts from previous work on TV commercials (See~\cite{shabbir2014deconstructing} for detailed reviews since 1950s).
One possible reason, of course, is the difference of target markets.

\begin{figure*}[ht!]
 \begin{center}
 \includegraphics[width=\columnwidth]{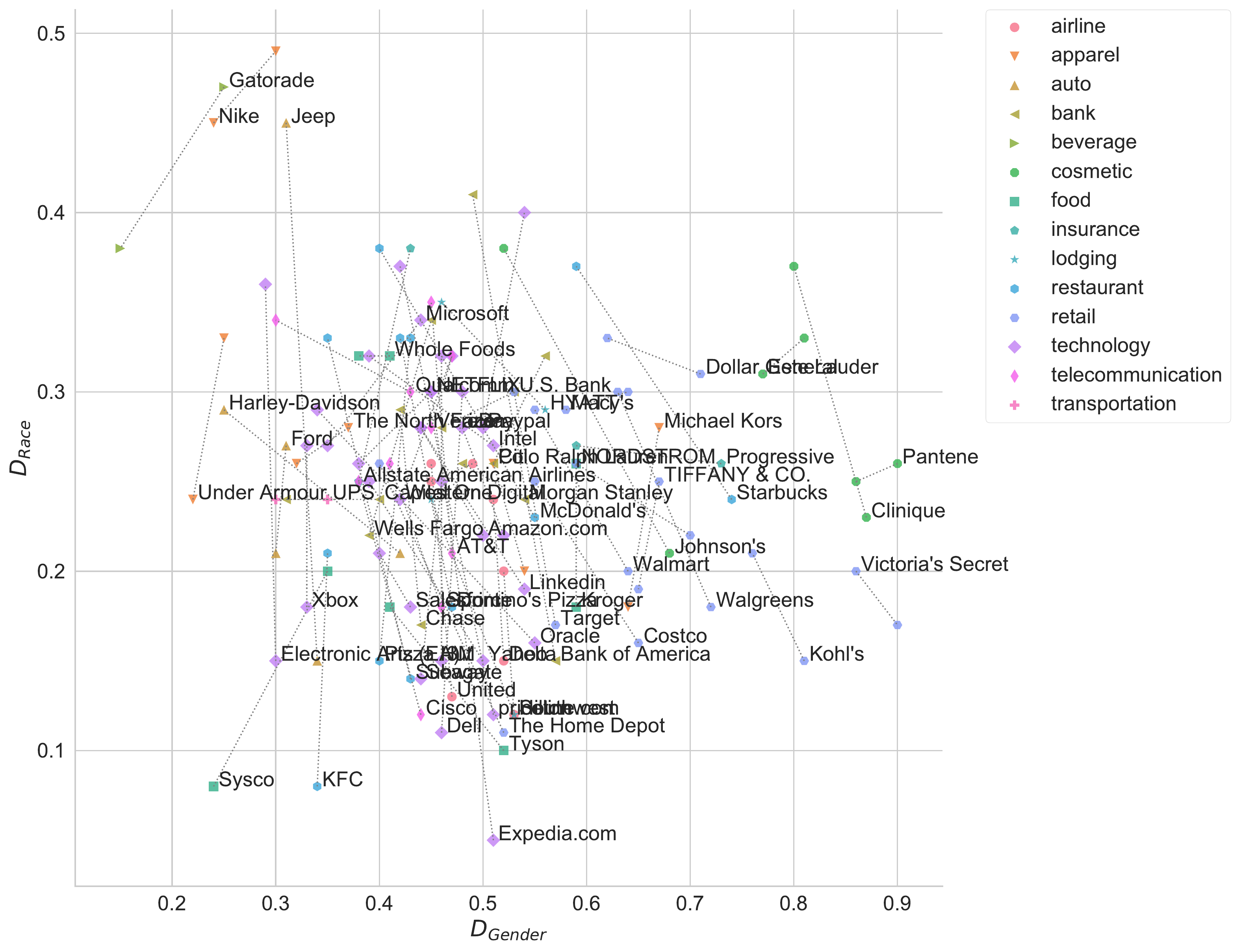}
 \caption{Gender and racial diversity in advertising images of brand on Instagram and Facebook. Each brand is represented by two markers linked by a dotted line, which are a marker with a label (Instagram) and that without a label (Facebook).} 
 \label{fig:diversity_map}
 \end{center}
\end{figure*}

To show the variances in frequencies of appearances of each demographic group in advertising images across the brands, we propose two metrics: gender diversity and racial diversity.
We define gender diversity of a brand $b$ as a ratio of appearances of female models in advertisements of the brand $b$ as follows:  $D^{b}_{gender}=\frac{N^b_{Female}}{N^b_{Female}+N^b_{Male}}$ 
where $N^b_{gender}$ is the number of appearances of models of that gender group in the advertisements of the brand $b$. 
Similarly, we define racial diversity of a brand $b$ as a ratio of appearances of non-white models in advertisements of the brand $b$ as follows:
$D^b_{race}=\frac{N^b_{Asian}+N^b_{Black}}{N^b_{Asian}+N^b_{Black}+N^b_{White}}$ 
where $N^b_{race}$ is the number of appearances of models of that racial group in the advertisements of the brand $b$. 
While the entropy is often used to measure the diversity, we simply use non-white ratio because (1) it has been widely used for racial diversity studies~\cite{shabbir2014deconstructing}, and (2) it is easy to interpret.

Figure~\ref{fig:diversity_map} shows the gender and racial diversity of the 73 brands on their advertising images on social media. The diversity values are separately computed from the advertisements on Instagram and Facebook.
Each brand is thus represented as two markers (one for Instagram and the other for Facebook) connected by a dotted line. To distinguish the two markers of one brand, we place a brand label at the marker for Instagram only.

We find some interesting trends in Figure~\ref{fig:diversity_map}. First, there is a huge variation in both diversity measures among the brands. For example, $D^{Expedia.com@Instagram}_{race}$= 0.05, which means that more than 90\% of models who appeared in their advertising images on Instagram are \white, while $D^{Nike@Facebook}_{race}$= 0.49, which means that around 50\% of models who appeared in their advertising images on Facebook are \asian~or \black. 
Second, by comparing the distributions of $D^{b}_{gender}$ and $D^{b}_{race}$, we can say that racial diversity should be more carefully considered. 
On average, $D^{b}_{gender}$ is 0.471 on Facebook and 0.505 on Instagram, but $D^{b}_{race}$ is 0.281 on Facebook and 0.219 on Instagram.

While the small sample size per sector does not allow us to conduct a statistical test for the sector differences, we observe that some brands show similar patterns on Instagram and Facebook regarding $D_{gender}$. 
The top three sectors are cosmetic, retail, and insurance, and the bottom three sectors are auto, beverage, and transportation on both platforms. Such strong patterns do not emerge in $D_{race}$, but beverage is the sector with the highest $D_{race}$ and food falls in bottom three sectors on both platforms.

\section{Diversity in Cross-Sex Interactions}

When multiple models appear in an advertisement, it delivers cues that these models are interacting, thus providing a better integration of models in advertising~\cite{dominick1970three,elliott1995differences}. 
In particular, advertising images in which models of different genders appear together, which are called cross-sex interactions, bring two specific situational contexts. 
One is a romantic context. Cross-sex interactions in advertisements fulfill romantic fantasies~\cite{coltrane2000perpetuation}, making viewers fall into the emotional experience and pay less attention to the sales pitch behind it~\cite{illouz1997consuming}. 
This means that models appearing in cross-sex interactions are likely to have some attractive qualities, such as beauty or masculinity, that can fulfill viewers' romantic fantasies. 
In this sense, Coltrane and Messineo~\cite{coltrane2000perpetuation} reveal the stereotyped White romantic fulfillment by reporting the prevalence of Whites, particularly females, in cross-sex interactions in TV commercials.
The other is a family context. Cross-sex interactions also imply another common setting in advertisements, which is a family. The happy family scene is one of the well crafted moments or fantasies that resonate with viewers~\cite{taylor1989prime}.  
In other words, models depicted in this setting signify family relationships to viewers and deliver cues to intimate relationships~\cite{coltrane2000perpetuation}.

\begin{figure}[ht!]
 \begin{center}
 \includegraphics[width=0.85\columnwidth]{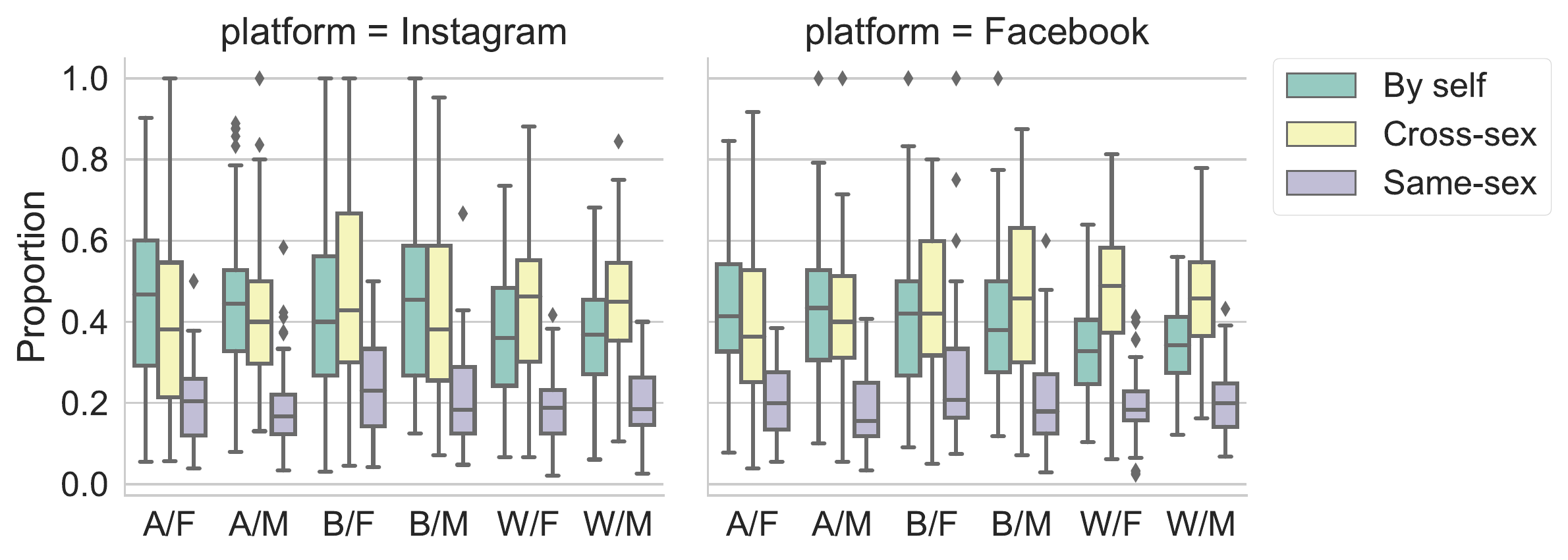}
 \caption{Proportion of each demographic group in cross-sex interactions, same-sex interaction, or by the self. A/, B/, and W/ stand for \asian, \black, and \white, and /F and /M stand for \female~and \male, respectively.} 
 \label{fig:each_dg_prop_in_cross_sex}
 \end{center}
\end{figure}

Figure~\ref{fig:each_dg_prop_in_cross_sex} shows the proportion of each demographic group appearing in cross-sex interaction, same-sex interaction (either men or women only), and by the self (a single person).
From the figure, three interesting patterns emerge. 
First, long whiskers show that there is a considerable variation of depictions of each demographic group. For example, in the advertising images posted by Victoria's Secret on Facebook, only 3.8\% of \af~models ($N$=8) are depicted in cross-sex interactions, while 65.9\% of \af~models (137) appear in the form of a single person (by the self). 
In contrast, in the advertising images of Citi on Facebook, 74.3\% of \af~models (26) are depicted in cross-sex interactions. 
This huge difference in cross-sex interactions across the brands, in addition to Figure~\ref{fig:diversity_map}, warns that selective exposure (i.e., following on social media) to certain brands can make the diversity problem more complicated.

Second, for any demographic group on both platforms, we can see a consistent trend that same-sex interaction is underrepresented compared to other forms of depictions. 
The statistically significant difference between the proportion of the same-sex interactions from that of cross-sex interactions and by-self are confirmed by Kruskal-Wallis 1-way ANOVA test with Dunn's post-hoc test (all the adjusted $p$ $<$ 1.159e-06 from all the comparisons).  This indicates that same-sex interactions are less preferred in advertising. Rather, a single model or cross-sex interactions tend to more appear.

Third, from the above post-hoc test, we confirm that the difference between by the self and cross-sex interactions is significant for \wm~($p$=3.15e-02) on Instagram,  \wf~($p$=5.81e-05) on Facebook, and \wm~($p$=1.45e-05) on Facebook. 
The difference in other demographic groups is not statistically significant. 
In other words, \white~models are more engaged in cross-sex interactions, while \asian~and \black~models are not. 
This aligns with the preference toward Whites in cross-sex interactions that have been repeatedly observed from TV commercials~\cite{coltrane2000perpetuation}. 

\begin{figure}[!ht]
 \begin{center}
\subfigure[]{ \includegraphics[width=0.49\columnwidth]{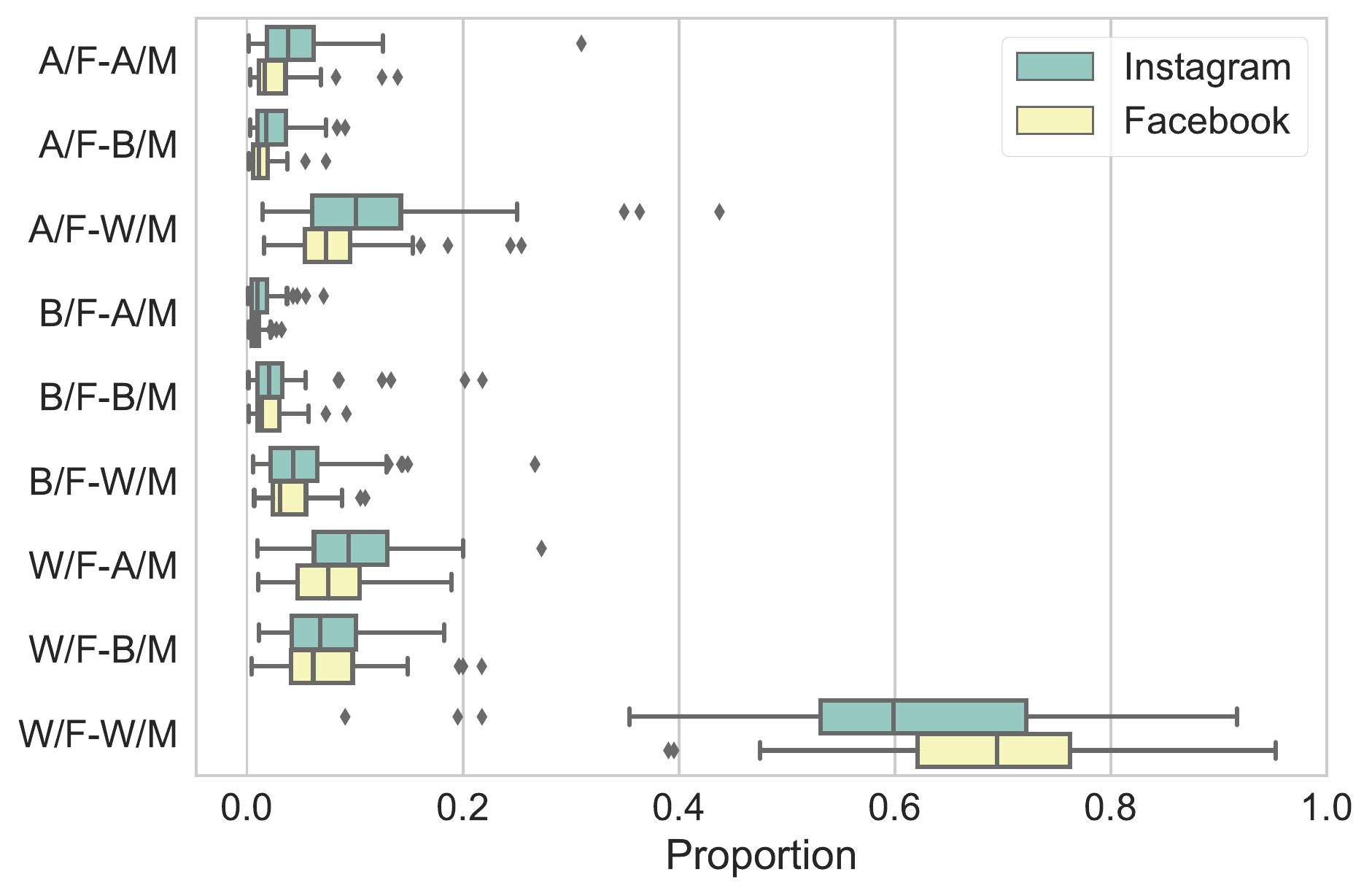}\label{fig:cross_sex_race_pair_prop}}
 \subfigure[]{ \includegraphics[width=0.49\columnwidth]{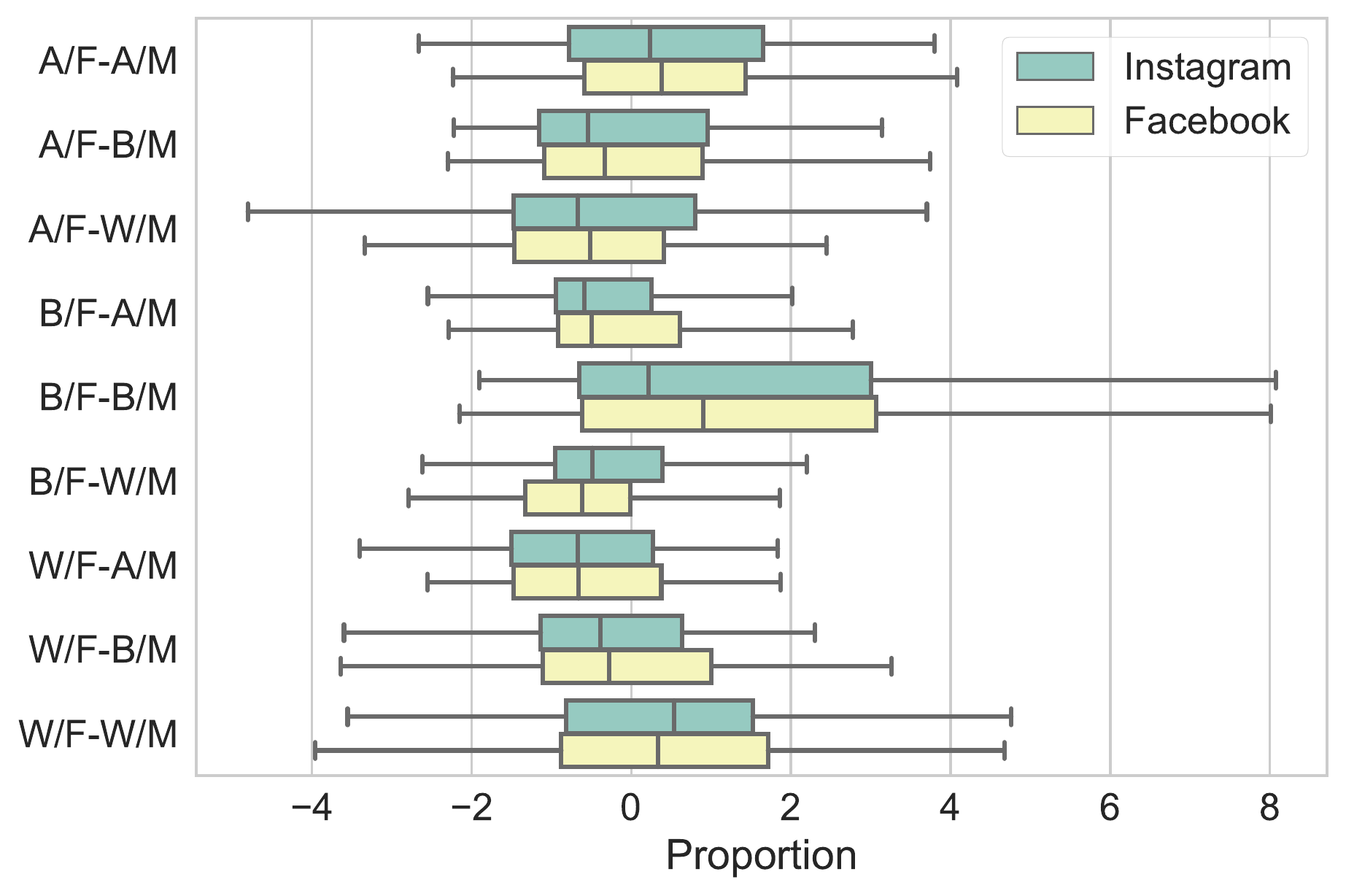}\label{fig:cross_sex_race_pair_Z}}
 \caption{(a) Proportions of each pair of different demographic groups in cross-sex interactions (b) Z-score of demographic group pairs in cross-sex interactions. Outliers are omitted for clarity} 
 \label{fig:cross_sex_race_pair}
 \end{center}
\end{figure}

Then, how is the bias toward the Whites presented at the level of the pairs in cross-sex interactions? 
To answer this question, we compute the proportion of all possible 9 combinations (3 races (Female side) $\times$ 3 races (Male side) = 9 combinations) of demographic groups from the advertising images that include cross-sex interactions.  
When there are more than two models in an advertisement, we take all the possible combinations of races between different genders into account. 
For example, if one Asian Female (A/F), one Asian Male (A/M), and one Black Male (B/M) models appear in an advertising image, we find two cross-sex interactions, which are (A/F-A/M) and (A/F-B/M).

Figure~\ref{fig:cross_sex_race_pair_prop} shows proportions of each pair of demographic groups in cross-sex interactions. 
To capture the variances between brands, we compute the proportions of each demographic group pair at the brand-level. 
From the figure, we observe that \wf~and \wm~models dominantly appear together. 
On average, 60.3\% and 68.9\% of cross-sex interactions in the advertising images are with \wf~and \wm~models on Instagram and Facebook, respectively. 
It is also noticeable that pairs with either \wf~or \wm~models are more frequent than those without any \white~model. 
However, this tendency does not mean that there is a systematic bias toward White models in pairing because, as we see in Figure~\ref{fig:simple_appearance}, \white~models simply outweigh other groups.
Thus, \white~models naturally have a higher chance to be shown in any cross-sex interactions than other groups due to its high number, even though there is no actual preference toward \wf~or \wm~models as a partner in cross-sex interaction.

To compute the corrected preference toward a pair of demographic groups in cross-sex interactions, we compare the actual proportion of each pair with the proportions of the corresponding pair from a null model. 
To build the null model, we randomly shuffle the placement of models across the advertising images while preserving the number of models of each gender in every advertising image. 
As we only ``shuffle'' the placement of existing models without addition or deletion of models, the total number of models in each demographic group stays the same. 
After shuffling a significant number of times (e.g., until each model is moved 10 times from one image to the other), we can get a null model that randomly locates models of each demographic group but preserves the number of the models of each gender in every advertising image. 
This null model shows the case that models appear in cross-sex interaction by chance. 
To avoid the errors of outliers, we build 1,000 null models and compute the average and the standard deviation of proportions of each pair of demographic groups.
Then, $Z$-score can be computed as follows:
$Z_{p_i} = \frac{N^{original}_{p_i}-avg(N^{rand}_{p_i})}{std(N^{rand}_{p_i})}$
where $p_i$ is a pair of certain demographic groups, $N^{original}_{p_i}$ is the number of pair $p_i$ observed in the original data, and $N^{rand}_{p_i}$ is the number of pair $p_i$ observed in a null model. 
The function of $avg(\cdot)$ is an average, and $std(\cdot)$ is a standard deviation. 
This $Z$-score shows how significant the observed frequency is compared to random. 
 
Figure~\ref{fig:cross_sex_race_pair_Z} shows Z-score of each pair of demographic groups engaged in cross-sex interactions on Instagram and Facebook. 
The most noticeable difference, compared to Figure~\ref{fig:cross_sex_race_pair_prop}, is that the preference toward pairing of the same race in cross-sex interactions becomes apparent. 
While there are variances among brands, the median Z-scores of pairs of \af~and \am~models, \kf~and \km, and \wf~and \wm~models in cross-sex interactions are positive on both Instagram and Facebook.
Although \white~models are frequently involved in cross-sex interactions as in Figure~\ref{fig:cross_sex_race_pair_prop}, interestingly, interracial interactions with \white~models actually have the negative $Z$-score, meaning that those pairs are less preferred in cross-sex interactions 
once taking the number of models in each demographic group into account.

The preference toward pairs of the same race in cross-sex interactions implies that there is an implicit opposition to interracial relationships in cross-sex interaction context. 
According to the 2010 Census~\cite{uscensus}, in the U.S., only 7\% of married couples are of different races, implying cross-race marriages are not common in society yet. 
Moreover, in a recent population-based survey experiment ($N$=2,035)~\cite{powell2017denial}, 39\% of respondents support the refusal of services to interracial couples. 
All these statistics prove that there is still a long way to go to the acceptance of interracial relationships, and advertising images reflect such reality to some extent.

\section{Demographic Groups with More (Less) Smiles}  

Based on the premise that non-verbal communications in human interaction are as important as verbal communications~\cite{knapp2013nonverbal}, there have been numerous attempts to understand facial expressions in advertisements~\cite{berg2015spreading,rodgers2007stereotypical}.
Since typical advertisements are made to capture attention, arouse interest, and lead to action~\cite{goddard2002language}, the most common facial expression of models is the \textit{smile} among ten basic facial expressions~\cite{ekman1976measuring}, and its positive impact on consumer attitude has been reported~\cite{berg2015spreading}.  
While there are some nuanced differences in emotions expressed by the smiles between females and males~\cite{vigil2009socio}, the smiling expression is quickly assessed with a high level of agreement and is more closely linked to happiness than other facial expressions to other emotions~\cite{wallbott1991recognition,lau1982effect}. 
Particularly, as advertisements usually deliver the positive side of products or services, the fact that happy faces accelerate the cognitive processing of positive words~\cite{stenberg1998judging} highlights the importance of smiling of the models in  advertisements.  
Based on this line of research, we examine how differently each demographic group expresses smiles in advertising images.

\begin{figure}[!ht]
 \begin{center}
 \includegraphics[width=0.85\columnwidth]{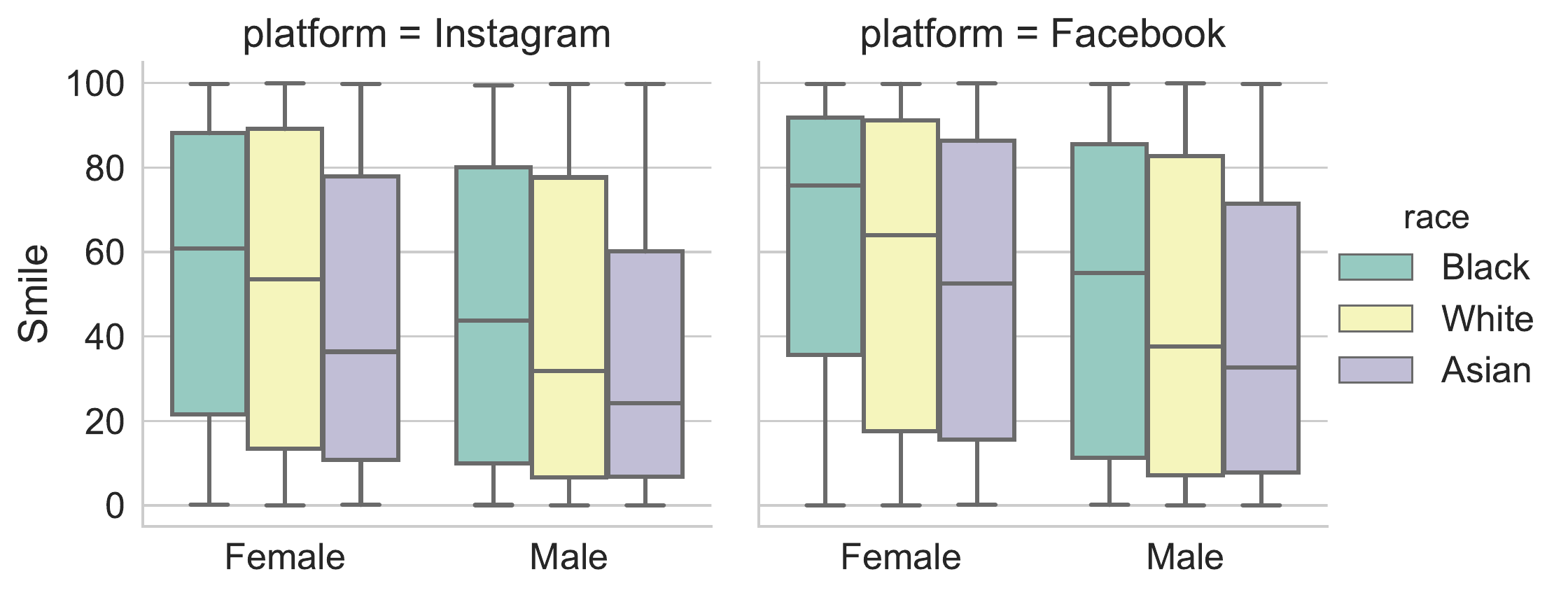}
 \caption{Smiling level of each demographic group} 
 \label{fig:smile_gender_diff}
 \end{center}
\end{figure}

Figure~\ref{fig:smile_gender_diff} shows the box plots of smiling levels (the degree of smile) of each demographic group. The smiling level ranges from 0 (no smile) to 100 (full smile) estimated by Face++. 
First, we find that \female~models smile more than \male~models where the mean smiling levels for \female~and \male~models are 55.6 ad.nd 42.8, respectively. 
The result is also statistically significant, confirmed by the independent t-test (Levene's test for equality of variance confirms that two sets have approximately equal variance (W=1.044, $p$=0.309), $t$=6.696, $p$ $<$ 0.001). 
The effect size, measured by Cohen's $d$, is 1.047, showing that there is a large effect of gender on smiling levels. 
This finding is consistent with previous studies that women smile more than men in various media~\cite{rodgers2007stereotypical,kwak2016revealing,lafrance2003contingent}. 
Of the same gender, differences between the races are also statistically significant. Any pair of different races of the same gender show significantly different levels of the smiles, confirmed by Kruskal-Wallis 1-way ANOVA test and the following Dunn's post-hoc test; all the $p$-values for the comparison of the different races are less than 3.30e-03.
Additionally, long boxes and whiskers reaching zero and one indicate that there exists a huge variance of the smiling levels even within one demographic group. 

\begin{figure}[!ht]
 \begin{center}
 \includegraphics[width=0.85\columnwidth]{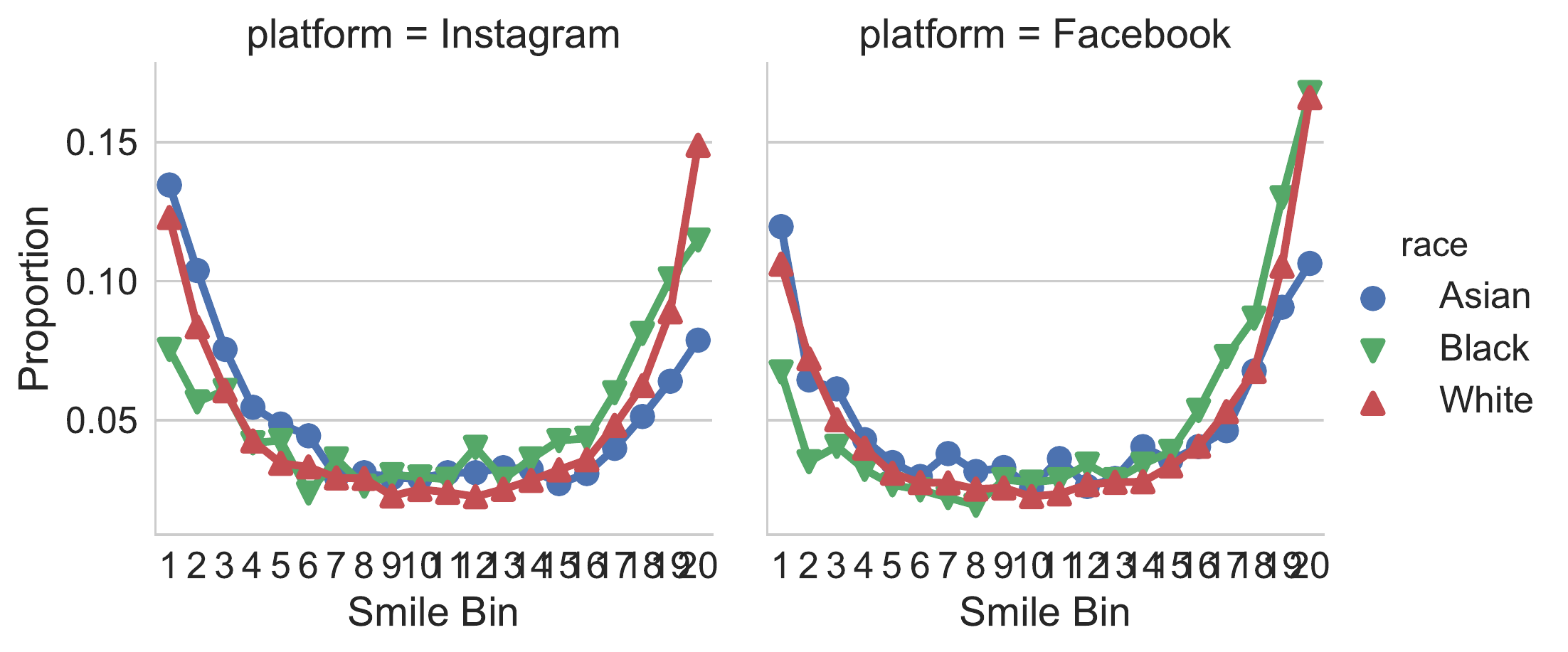}
 \includegraphics[width=0.85\columnwidth]{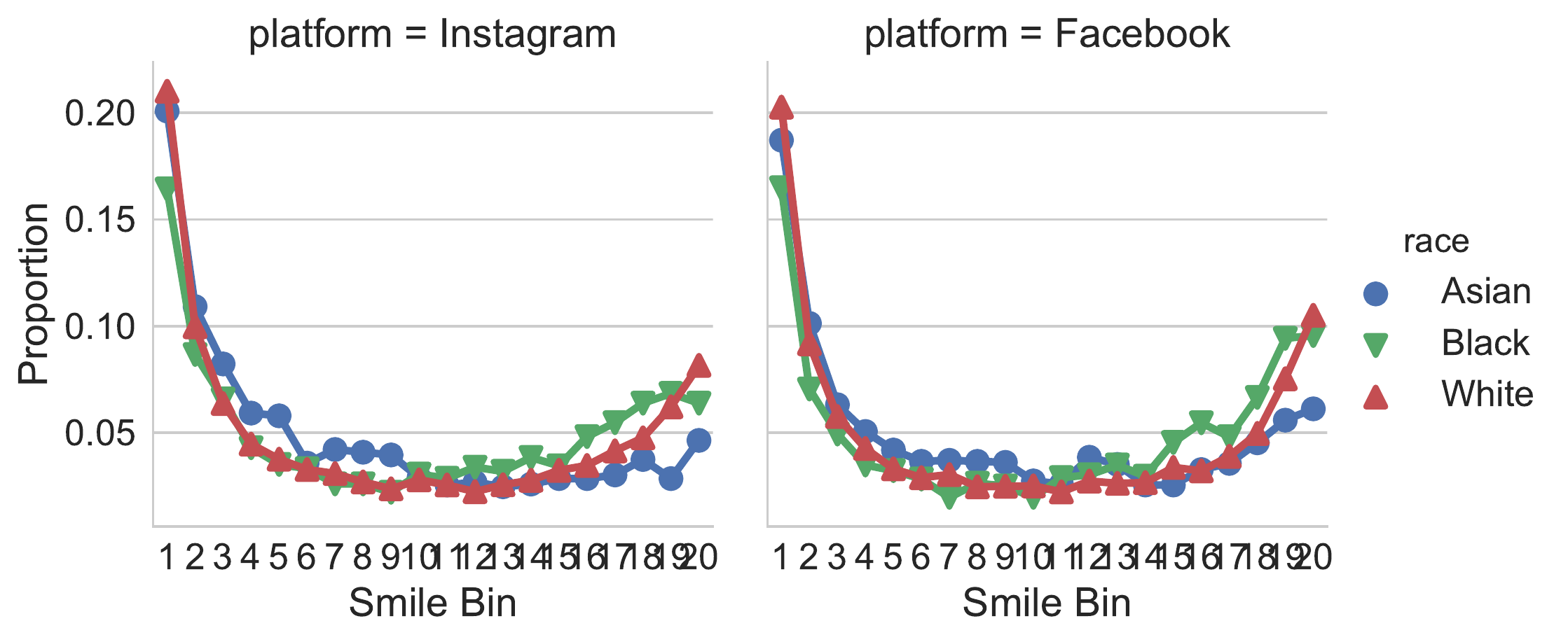}
 \caption{Proportions of each race in different smiling levels (Top: \female, Bottom: \male)} 
 \label{fig:prop_race_smile_level_male}
 \end{center}
\end{figure}

Figure \ref{fig:prop_race_smile_level_male} shows the distribution of the smiling levels of each race group of females (top) and males (bottom), respectively. 
Here we quantize the smiling levels into 20 bins. 
From all figures, U-shaped (balanced and unbalanced) curves consistently emerge; every demographic group has a larger proportion of non-smile faces (leftmost bin) or full-smile faces (rightmost bin) than faces with other smiling levels.
Also, \male~models are more often depicted with non-smile than full-smile, while \female~models are more often depicted with full-smile than non-smile except \af~models. 
Together with Figure~\ref{fig:smile_gender_diff}, these results resonate with previous work~\cite{vigil2009socio} that social constraints imposed on females who live in a male-biased society make themselves to rely on ``behaviors designed to advertise their trustworthiness through higher levels of submissive displays,'' and smiling works in that way, as a cue for the trustworthiness of women~\cite{vazire2009smiling}.

\section{Discussions and Conclusions}

In this work, we proposed an automated way to examine gender and racial diversity in online advertisements using modern image analysis techniques. We demonstrated its feasibility through a large-scale analysis of 85,957 advertising images of 73 international brands on Instagram and Facebook.

The quantitative analysis has shown several interesting findings. First, we observed huge variations of gender and racial diversity in advertising images across the brands, as those images are made to fit their models and representations in social norms, cultural values, target market, and social media audiences~\cite{cohen2004advertisements}. 
As brands have their own target audiences, it is reasonable to have even very skewed model representations (e.g., Victoria's Secret uses more than 90\% \female~models).
Second, nevertheless, we found consistent trends that \white~models outnumber other races. 
Also, we observed \asian~models are overrepresented which could be due to brands' efforts to appeal to a huge Asian market.
By contrast, \black~models are quite underrepresented. 
In contrast to a recent study of increasing appearances of Black models in advertisements in broadcast and print media~\cite{shabbir2014deconstructing}, their presence on advertising images in social media still stays low. 
Third, there are some differences in gender and racial diversity between Instagram and Facebook.  
The observed differences, higher percentages of racial diversity on Facebook and higher gender diversity on Instagram, are well aligned with their userbase~\cite{an2018diversity}. 
Fourth, \white~models are more engaged in cross-sex interactions, as previous literature reported~\cite{coltrane2000perpetuation}. 
In addition, by comparing the actual data with random null models, we revealed the preference toward the same race pairs in cross-sex interactions. In other words, inter-racial cross-sex interactions are underrepresented.
Finally, we captured gender bias that \female~models smile more than \male~models, which is consistent with previous studies in various media~\cite{rodgers2007stereotypical,kwak2016revealing,lafrance2003contingent}.

Our work is not without limitations. First, the use of Face detection for inferring gender and race may have inherent biases~\cite{buolamwini2018gender}.
Second, we consider global top brands and do not take cultural differences into account. Since there could exist huge variances in gender and racial diversity in advertising images across regions and cultures, follow-up studies with ore data are required for generalization. Third, we were unable to analyze the nuanced portrayals of models. For example, an advertisement with female chefs and male engineers might have a high gender diversity but reinforce gendered stereotypes. Recently, such `stereotyped gender roles' have been actively studied~\cite{kay2015unequal,bolukbasi2016man}.
For future work, we aim to develop new computational methods that can capture more sophisticated representations of models in advertisements.

Notwithstanding the limitation, this work contributes to the stream of research on gender and racial diversity in advertisements by introducing the fully automated computational tools on advertisement studies and conducting a large-scale analysis. 
The entire process, which is collecting advertising images, labeling faces, and computing the diversity scores, is automated and will become more accurate as the image recognition technology improves.
It is thus possible to monitor the diversity efforts of brands with extremely low cost and near real-time. 
This approach is a substantial advantage compared to the required time and resource in previous research, such as collecting TV commercials or magazine advertisements, labeling faces by educated workers, and so on. 
While there is room for improvement particularly in capturing a nuanced representation of models, our effort is the first step to build an automated pipeline for tracking gender and racial diversity in the advertisements and raising public awareness toward the right depiction of an ethnic minority in the media.

\balance
%
%
%
\bibliographystyle{splncs04}
\bibliography{bib-brand-diversity}

\begin{thebibliography}{10}
\providecommand{\url}[1]{\texttt{#1}}
\providecommand{\urlprefix}{URL }
\providecommand{\doi}[1]{https://doi.org/#1}

\bibitem{interbrand}
Best global brands 2016.
  \url{https://www.interbrand.com/best-brands/best-global-brands/2016/},
  accessed: 2019-04-16

\bibitem{facepp}
Face++. \url{https://www.faceplusplus.com}, accessed: 2019-04-16

\bibitem{hornbach}
German diy chain's `racist' advert provokes anger in south korea.
  \url{https://www.theguardian.com/world/2019/mar/28/german-diy-chain-hornbach-racist-advert-provokes-anger-south-korea},
  accessed: 2019-04-16

\bibitem{brandirectory}
Global 500 2016 | rankings | brandirectory.
  \url{https://brandirectory.com/rankings/global-500-2016}, accessed:
  2019-04-16

\bibitem{uscensus}
Households and families: 2010.
  \url{https://www.census.gov/prod/cen2010/briefs/c2010br-14.pdf}, accessed:
  2019-04-16

\bibitem{an2016greysanatomy}
An, J., Weber, I.: \#greysanatomy vs. \#yankees: Demographics and hashtag use
  on {T}witter. In: ICWSM (2016),
  \url{https://www.aaai.org/ocs/index.php/ICWSM/ICWSM16/paper/view/13021}

\bibitem{an2018diversity}
An, J., Weber, I.: Diversity in online advertising: A case study of 69 brands
  on social media. In: International Conference on Social Informatics. pp.
  38--53. Springer (2018)

\bibitem{bailey2006year}
Bailey, A.A.: A year in the life of the african-american male in advertising: A
  content analysis. Journal of Advertising  \textbf{35}(1),  83--104 (2006)

\bibitem{bandura1969social}
Bandura, A.: Social-learning theory of identificatory processes. Handbook of
  socialization theory and research  \textbf{213}, ~262 (1969)

\bibitem{berg2015spreading}
Berg, H., S{\"o}derlund, M., Lindstr{\"o}m, A.: Spreading joy: examining the
  effects of smiling models on consumer joy and attitudes. Journal of Consumer
  Marketing  \textbf{32}(6),  459--469 (2015)

\bibitem{bolukbasi2016man}
Bolukbasi, T., Chang, K.W., Zou, J.Y., Saligrama, V., Kalai, A.T.: Man is to
  computer programmer as woman is to homemaker? debiasing word embeddings. In:
  Advances in neural information processing systems. pp. 4349--4357 (2016)

\bibitem{bowen1997minority}
Bowen, L., Schmid, J.: Minority presence and portrayal in mainstream magazine
  advertising: An update. Journalism \& Mass Communication Quarterly
  \textbf{74}(1),  134--146 (1997)

\bibitem{buolamwini2018gender}
Buolamwini, J., Gebru, T.: Gender shades: Intersectional accuracy disparities
  in commercial gender classification. In: Conference on fairness,
  accountability and transparency. pp. 77--91 (2018)

\bibitem{choudhury1974black}
Choudhury, P.K., Schmid, L.S.: Black models in advertising to blacks. Journal
  of Advertising Research  \textbf{14}(3),  19--22 (1974)

\bibitem{cohen2004advertisements}
Cohen-Eliya, M., Hammer, Y.: Advertisements, stereotypes, and freedom of
  expression. Journal of Social Philosophy  \textbf{35}(2),  165--187 (2004)

\bibitem{coltrane2000perpetuation}
Coltrane, S., Messineo, M.: The perpetuation of subtle prejudice: Race and
  gender imagery in 1990s television advertising. Sex roles  \textbf{42}(5),
  363--389 (2000)

\bibitem{cox1969changes}
Cox, K.K.: Changes in stereotyping of negroes and whites in magazine
  advertisements. The Public Opinion Quarterly  \textbf{33}(4),  603--606
  (1969)

\bibitem{dominick1970three}
Dominick, J.R., Greenberg, B.S.: Three seasons of blacks on television. Journal
  of Advertising Research  \textbf{26},  169--173 (1970)

\bibitem{ekman1976measuring}
Ekman, P., Friesen, W.V.: Measuring facial movement. Environmental psychology
  and nonverbal behavior  \textbf{1}(1),  56--75 (1976)

\bibitem{elliott1995differences}
Elliott, M.T.: Differences in the portrayal of blacks: A content analysis of
  general media versus culturally-targeted commercials. Journal of Current
  Issues \& Research in Advertising  \textbf{17}(1),  75--86 (1995)

\bibitem{gerbner2002growing}
Gerbner, G., Gross, L., Morgan, M., Signorielli, N., Shanahan, J.: Growing up
  with television: Cultivation processes. Media effects: Advances in theory and
  research  \textbf{2},  43--67 (2002)

\bibitem{goddard2002language}
Goddard, A.: The language of advertising: written texts. Psychology Press
  (2002)

\bibitem{greenwald1998measuring}
Greenwald, A.G., McGhee, D.E., Schwartz, J.L.: Measuring individual differences
  in implicit cognition: the implicit association test. Journal of personality
  and social psychology  \textbf{74}(6), ~1464 (1998)

\bibitem{hanna2011we}
Hanna, R., Rohm, A., Crittenden, V.L.: We’re all connected: The power of the
  social media ecosystem. Business horizons  \textbf{54}(3),  265--273 (2011)

\bibitem{illouz1997consuming}
Illouz, E.: Consuming the romantic utopia: Love and the cultural contradictions
  of capitalism. Univ of California Press (1997)

\bibitem{jacobs2003race}
Jacobs~Henderson, J., Baldasty, G.J.: Race, advertising, and prime-time
  television. Howard Journal of Communication  \textbf{14}(2),  97--112 (2003)

\bibitem{jung2018facepp}
Jung, S., An, J., Kwak, H., Salminen, J., Jansen, B.: Assessing the accuracy of
  four popular face recognition tools for inferring gender, age, and race. In:
  ICWSM (2018)

\bibitem{katz1988racial}
Katz, I., Hass, R.G.: Racial ambivalence and american value conflict:
  Correlational and priming studies of dual cognitive structures. Journal of
  personality and social psychology  \textbf{55}(6), ~893 (1988)

\bibitem{kay2015unequal}
Kay, M., Matuszek, C., Munson, S.A.: Unequal representation and gender
  stereotypes in image search results for occupations. In: Proceedings of the
  33rd Annual ACM Conference on Human Factors in Computing Systems. pp.
  3819--3828. ACM (2015)

\bibitem{knapp2013nonverbal}
Knapp, M.L., Hall, J.A., Horgan, T.G.: Nonverbal communication in human
  interaction. Cengage Learning (2013)

\bibitem{kwak2016revealing}
Kwak, H., An, J.: Revealing the hidden patterns of news photos: Analysis of
  millions of news photos using gdelt and deep learning-based vision apis. In:
  ICWSM workshop on news and public opinion (NECO) (2016)

\bibitem{kwak2018we}
Kwak, H., An, J., Salminen, J., Jung, S.G., Jansen, B.J.: What we read, what we
  search: Media attention and public attention among 193 countries. In:
  Proceedings of the 2018 World Wide Web Conference on World Wide Web. pp.
  893--902. International World Wide Web Conferences Steering Committee (2018)

\bibitem{lafrance2003contingent}
LaFrance, M., Hecht, M.A., Paluck, E.L.: The contingent smile: a meta-analysis
  of sex differences in smiling. Psychological bulletin  \textbf{129}(2), ~305
  (2003)

\bibitem{lau1982effect}
Lau, S.: The effect of smiling on person perception. The Journal of Social
  Psychology  \textbf{117}(1),  63--67 (1982)

\bibitem{lee2005portrayal}
Lee, K.Y., Joo, S.H.: The portrayal of asian americans in mainstream magazine
  ads: An update. Journalism \& Mass Communication Quarterly  \textbf{82}(3),
  654--671 (2005)

\bibitem{mastro2003representations}
Mastro, D.E., Stern, S.R.: Representations of race in television commercials: A
  content analysis of prime-time advertising. Journal of Broadcasting \&
  Electronic Media  \textbf{47}(4),  638--647 (2003)

\bibitem{mccullick2003butches}
McCullick, B., Belcher, D., Hardin, B., Hardin, M.: Butches, bullies and
  buffoons: Images of physical education teachers in the movies. Sport,
  Education and Society  \textbf{8}(1),  3--16 (2003)

\bibitem{mcquarrie2008differentiating}
McQuarrie, E.F.: Differentiating the pictorial element in advertising. Visual
  Marketing: From Attention to Action. New York: Erlbaum pp. 91--112 (2008)

\bibitem{millard2006stereotypes}
Millard, J.E., Grant, P.R.: The stereotypes of black and white women in fashion
  magazine photographs: The pose of the model and the impression she creates.
  Sex Roles  \textbf{54}(9-10),  659--673 (2006)

\bibitem{paek2003racial}
Paek, H.J., Shah, H.: Racial ideology, model minorities, and the" not-so-silent
  partner:" stereotyping of asian americans in us magazine advertising. Howard
  Journal of Communication  \textbf{14}(4),  225--243 (2003)

\bibitem{plous1997racial}
Plous, S., Neptune, D.: Racial and gender biases in magazine advertising: A
  content-analytic study. Psychology of women quarterly  \textbf{21}(4),
  627--644 (1997)

\bibitem{powell2017denial}
Powell, B., Schnabel, L., Apgar, L.: Denial of service to same-sex and
  interracial couples: Evidence from a national survey experiment. Science
  advances  \textbf{3}(12),  eaao5834 (2017)

\bibitem{reis2017demographics}
Reis, J.C., Kwak, H., An, J., Messias, J., Benevenuto, F.: Demographics of news
  sharing in the u.s. twittersphere. In: Proceedings of the 28th ACM Conference
  on Hypertext and Social Media. pp. 195--204. HT '17, ACM (2017).
  \doi{10.1145/3078714.3078734},
  \url{http://doi.acm.org/10.1145/3078714.3078734}

\bibitem{richeson2005brief}
Richeson, J.A., Shelton, J.N.: Brief report: Thin slices of racial bias.
  Journal of Nonverbal Behavior  \textbf{29}(1),  75--86 (2005)

\bibitem{rodgers2007stereotypical}
Rodgers, S., Kenix, L.J., Thorson, E.: Stereotypical portrayals of emotionality
  in news photos. Mass Communication \& Society  \textbf{10}(1),  119--138
  (2007)

\bibitem{shabbir2014deconstructing}
Shabbir, H.A., Hyman, M.R., Reast, J., Palihawadana, D.: Deconstructing subtle
  racist imagery in television ads. Journal of business ethics
  \textbf{123}(3),  421--436 (2014)

\bibitem{shuey1953stereotyping}
Shuey, A.M., King, N., Griffith, B.: Stereotyping of negroes and whites: An
  analysis of magazine pictures. Public Opinion Quarterly pp. 281--287 (1953)

\bibitem{skill1990family}
Skill, T., Wallace, S.: Family interactions on primetime television: A
  descriptive analysis of assertive power interactions. Journal of Broadcasting
  \& Electronic Media  \textbf{34}(3),  243--262 (1990)

\bibitem{stenberg1998judging}
Stenberg, G., Wiking, S., Dahl, M.: Judging words at face value: Interference
  in a word processing task reveals automatic processing of affective facial
  expressions. Cognition \& Emotion  \textbf{12}(6),  755--782 (1998)

\bibitem{stevenson2007six}
Stevenson, T.H.: A six-decade study of the portrayal of african americans in
  business print media: Trailing, mirroring, or shaping social change? Journal
  of Current Issues \& Research in Advertising  \textbf{29}(1),  1--14 (2007)

\bibitem{swani2014should}
Swani, K., Brown, B.P., Milne, G.R.: Should tweets differ for {B2B} and {B2C}?
  an analysis of {F}ortune 500 companies' {T}witter communications. Industrial
  Marketing Management  \textbf{43}(5),  873--881 (2014)

\bibitem{taylor1994not}
Taylor, C.R., Lee, J.Y.: Not in vogue: Portrayals of asian americans in
  magazine advertising. Journal of Public Policy \& Marketing pp. 239--245
  (1994)

\bibitem{taylor1997asian}
Taylor, C.R., Stern, B.B.: Asian-americans: Television advertising and the
  “model minority” stereotype. Journal of advertising  \textbf{26}(2),
  47--61 (1997)

\bibitem{taylor1989prime}
Taylor, E.: Prime-time families: television culture in post-war {A}merica. Univ
  of California Press (1989)

\bibitem{trebbe2011ethnic}
Trebbe, J., Schoenhagen, P.: Ethnic minorities in the mass media: How migrants
  perceive their representation in swiss public television. Journal of
  International Migration and Integration  \textbf{12}(4),  411--428 (2011)

\bibitem{vazire2009smiling}
Vazire, S., Naumann, L.P., Rentfrow, P.J., Gosling, S.D.: Smiling reflects
  different emotions in men and women. Behavioral and brain sciences
  \textbf{32}(5),  403--405 (2009)

\bibitem{vigil2009socio}
Vigil, J.M.: A socio-relational framework of sex differences in the expression
  of emotion. Behavioral and Brain Sciences  \textbf{32}(5),  375--390 (2009)

\bibitem{wallbott1991recognition}
Wallbott, H.G.: Recognition of emotion from facial expression via imitation?
  some indirect evidence for an old theory. British Journal of Social
  Psychology  \textbf{30}(3),  207--219 (1991)

\bibitem{zagheni2014inferring}
Zagheni, E., Garimella, V.R.K., Weber, I., et~al.: Inferring international and
  internal migration patterns from {T}witter data. In: Proceedings of WWW
  Companion (2014)

\bibitem{zinkhan1990use}
Zinkhan, G.M., Quails, W.J., Biswas, A.: The use of blacks in magazine and
  television advertising: 1946 to 1986. Journalism Quarterly  \textbf{67}(3),
  547--553 (1990)

\end{thebibliography}

\end{document}